\documentstyle[12pt,aaspp4]{article} 
\def\tempest%
{\begin{array}{ccc} 
1 & 1 & 1 \\ 
1 & 1 & 1 \\ 
4 & 3 & 8 
\end{array}} 

\def\bOmega{{\hbox{$\Omega\hskip-8.0pt\Omega$}}}
\def\btheta{{\hbox{$\theta\hskip-6.3pt\theta$}}}

\def\bmu{\hbox{$\mu\hskip-7.5pt\mu$}}
\def\kms{{\rm km}\,{\rm s}^{-1}}
\def\br{{\bf r}}
\def\bn{{\bf n}}
\def\bV{{\bf V}}
\def\bv{{\bf v}}
\def\bu{{\bf u}}
\def\sn{{\rm sn}}
\def\dlmc{{d_{\rm LMC}}}
\begin{document}

\title{A New Kinematic Distance Estimator to the 
LMC}
\author 
{Andrew Gould}
\affil{Ohio State University, Department of Astronomy, Columbus, OH 43210} 
\affil{E-mail: gould@astronomy.ohio-state.edu} 
\begin{abstract} 

	The distance to the Large Magellanic Cloud (LMC) can be directly
determined by measuring three of its properties, its radial-velocity
field, its mean proper motion, and the position angle $\phi_{\rm ph}$
of its photometric line of nodes.  Statistical errors of $\sim 2\%$ are
feasible based on proper motions obtained with any of several proposed
astrometry satellites, the first possibility being the Full-Sky Astrometric
Mapping Explorer (FAME).  The largest source of systematic error is
likely to be in the determination of $\phi_{\rm ph}$.  I suggest two 
independent methods to measure $\phi_{\rm ph}$, one based on counts of
clump giants and the other on photometry of clump giants.  I briefly
discuss a variety of methods to test for other sources of systematic
errors.

\end{abstract} 
\newpage

\section{Introduction} 

	The distance to the Large Magellanic Cloud (LMC) plays a crucial
role in the extragalactic distance scale.  The relation between log-period
and apparent magnitude of LMC Cepheids is quite well determined.  If
an LMC distance $\dlmc$ and mean LMC-Cepheid reddening are assumed, then
the Cepheid period-luminosity relation is effectively calibrated.  
The distance to external galaxies harboring Cepheids can then be determined
by comparing their observed fluxes to those of LMC Cepheids at the same
period, and by taking account of the differences in reddening which are
determined from the differences in color between the target Cepheids and
those in the LMC.  A variety of secondary distance indicators have been
calibrated in this fashion.  While a decade ago, the Hubble constant
$H_0$ derived from these measurements ranged over a factor of two depending
strongly on both the author and the method, a major observing campaign
with the Hubble Space Telescope (HST) has dramatically narrowed this conflict.
 For example, 
Saha et al.\ (1999) recently find 
$H_0 \dlmc/50\,\rm kpc=60\pm 2\,(\rm internal)\,
\rm km\,s^{-1}\,Mpc^{-1}$ 
while 
Madore et al.\ 1999 find
$H_0 \dlmc/50\,\rm kpc=72\pm 3\,(\rm random)\,\pm 5\, (systematic)
\,\rm km\,s^{-1}\,Mpc^{-1}$.

	By constrast, the disagreements over $\dlmc$ have not
narrowed at all over the past decade.  The primary methods for measuring
$\dlmc$ use ``standard candles'', objects whose luminosity is presumed
to be fixed or to depend only on distance-independent observables such
as period, metallicity, etc.  Their absolute magnitudes must be calibrated
locally.  Two major standard candles that have been used to measure $\dlmc$
are Cepheids and RR Lyrae stars.  Three recent determinations, all from 
Hipparcos-based calibrations of these standard candles, illustrate the range
of $\dlmc$ estimates: Feast \& Catchpole (1997) find $\dlmc=55.0\pm 2.5\,$kpc
based on trigonometric parallaxes of Cepheids; 
Gould \& Popowski (1998) find $\dlmc=45.1\pm 2.7\,$kpc based on statistical
parallax of RR Lyraes [and assuming $V_0(RR,LMC)=18.98$,  Walker 1992];
and Gratton et al.\
(1997) finds $\dlmc=52.1\pm 1.7\,$kpc from an RR Lyrae calibration
based on fitting globular cluster main sequences to Hipparcos subdwarfs,
while Reid (1997) finds a slightly longer distance based on the same
technique.  At present it is not known if these discrepancies are due
to undetected systematic errors in the various techniques or to 
non-standardness
in one or more of the ``standard candles'', or both.  It is unlikely that
that the differences are merely statistical fluctuations.  For example,
Popowski \& Gould (1999) review a variety of methods for calibrating
RR Lyrae stars whose results disagree by substantially more than their
statistical errors.

	Of course, one would prefer to eliminate the distance ladder
altogether and simply obtain a direct measurement of $\dlmc$
There are two possible paths to a direct distance measurement: trigonometric
parallax and kinematic methods.  The parallax of the LMC is
$\pi \sim 20\,\mu$as.  The Space Interferometry Mission
(SIM) should be able to make individual astrometric measurements
accurate to $\sim 8\,\mu$as, and  could perhaps achieve
$\sigma_\pi\sim 2\,\mu$as given a sufficient number of observations.
This limit is set by the precision of the SIM ``grid-star'' solution.  Hence
it cannot be significantly improved upon by making measurements of 
several LMC stars, since these lie in the same field.  While such a
$\sim 10\%$ meausurement would certainly be of interest, it would not by 
itself clearly distinguish among the various competing distance estimates.

	Panageia et al.\ (1991) made the first kinematic measurement
of $\dlmc$ by comparing the light travel time accross the ring around
Supernova 1987A with its angular diameter as measured by HST.  They
found $\mu_\sn=18.55\pm 0.13$ where $\mu_\sn$ is the distance modulus of 
Supernova 1987A.  Gould (1995b) reanalyzed these data
and obtained  $\mu_\sn\leq 18.350\pm 0.035$.  Sonneborne et al.\ (1996)
rereduced the original light-curve data and found $\mu_\sn = 18.43\pm 0.10$.
Gould \& Uza (1998) then reanalyzed these rereduced data and obtained
$\mu_\sn\leq 18.372\pm 0.035$ if the ring were assumed circular, but
$\mu_\sn\leq 18.44\pm 0.05$ if it were assumed elliptical (as some
evidence suggests).  Finally, Panagia (1998), using the same data, but
arguing that the effective radius of the ring had grown between the time
of the light echo measurements and those of the angular size of the ring,
found $\mu_\sn=18.55\pm 0.05$.  In brief, there remains controversy over
the interpretion of the data at the $\sim 10\%$ level in distance.  Since
the event itself was unique and the measurements will never be repeated,
it seems unlikely that this conflict will be resolved to everyone's 
satisfaction.

	Here I propose to use the radial-velocity
gradient method to measure the distance to the LMC.  This method has been
used in the past to measure the distance to the Hyades (Detweiler et al.\ 1984;
Gunn et al.\ 1988) and the Pleiades (Narayanan \& Gould 1999).  When applied
to the LMC, the method has some unique characteristics relative to previous
applications.  This is in part because the LMC is a cold system supported 
primarily by rotation while the Hyades and Pleiades are supported by
pressure, and in part because the LMC is two orders of magnitude farther away.

	Consider a cold disk rotating at a projected angular rate $\Omega(R)$,
and moving with a systemic proper motion $\bmu$ (and so transverse velocity
$\bV_\perp=\bmu d$).  If $\mu\ll \Omega$, then the locus of extrema
in the radial velocity field will coincide with the photometric major axis.
That is, the kinematic and photometric
lines of nodes will be aligned.  However, the transverse velocity $\bV_\perp$
of any system induces a gradient in the radial velocities because the
radial vector that is dotted into the velocity to form the radial velocity
changes direction across the system.  Thus the
observed gradient $\nabla v_r$ will be displaced from the that due to internal
rotation alone by $\bV_\perp$.  If the direction of the photometric line
of nodes is known, and if $\bmu$ is
measured (so that the direction of $\bV_\perp$ is also known), then it
is straight forward to solve for the magnitude of $\bV_\perp$.  The
distance is then simply $d=V_\perp/\mu$.  The LMC is sufficiently close
and is moving sufficiently rapidly, that the kinematic line of nodes
is displaced from the photometric line of nodes by about $25^\circ$.
The interpretation of this displacement could be clouded by uncertainty
about how well the LMC conforms to the ideal of a flat axisymmetric system
that I use to model the data.  After I present the method and derive the
statistical uncertainties, I briefly discuss how the measurement
could be corrupted by systematic deviations from this ideal, and I 
indicate some methods to check for such systematic effects.

To illustrate the method, I will assume the use of astrometry data such
as would be obtained by the Full-Sky Astrometric Explorer (FAME), a proposed
Midex mission.  As I will discuss, the method could also be applied to 
data from SIM or the Global Astrometric Interferometer for Astrophysics (GAIA).

\section{The Method}

	Consider a stellar system whose physical size is small compared to its
distance $d$.  Let the space velocity of the system be $\bV$ and let
the space motion of an individual star in the sysetm be $\bv_i$.  I then
write
\begin{equation}
\bv_i = \bV + \bu_i + \delta\bv_i
\label{eqn:vdecomp}
\end{equation}
where $\bu_i$ is the mean internal systemic motion of the stars in the system 
(due, e.g., to rotation) at the projected position of star $i$, and
$\delta\bv_i$ is the peculiar motion of star $i$ relative to this systemic
motion.  The radial-velocities are therefore given by,
\begin{equation}
v_{r,i} = \bn_i\cdot \bV + u_{r,i} + \delta v_{r,i}
\label{eqn:radvel}
\end{equation}
where $\bn_i$ is the unit vector in the direction of star $i$,
$u_{r,i}=\bn_i\cdot \bu_i$, and $\delta v_{r,i}=\bn_i\cdot \delta \bv_i$.
I assume that the radial-velocity residuals $\delta v_{r,i}$
are randomly distributed with dispersion $\sigma_v$.  I also assume that
the internal systemic motion $\bu_i$ is known.  In fact, determining $\bu$
is not trivial, but I ignore this problem here and return to it in
\S\S\ 3.2, 3.3, and 4.
Then the radial-velocity gradient with respect to angular position
on the sky is given by
\begin{equation}
\nabla v_r  = \bV_\perp + \nabla u_r
\label{eqn:nabla}
\end{equation}
where $\bV_\perp = \bV - \bn_0 (\bn_0\cdot \bV)$ is the transverse velocity
of the center of the system, and $\bn_0$ is the direction vector pointing
to this center.  Since $\nabla v_r$ is a vector, the errors are properly 
described by
a covariance matrix, $c_{x y}$.  This is given by
\begin{equation}
c\equiv b^{-1}\qquad b_{kl} = {1\over \sigma_v^2}\biggl[
\sum_i \theta_{k,i}\theta_{l,i} - {1\over N_r}
\biggl(\sum_i\theta_{k,i}\biggr)\biggl(\sum_i\theta_{l,i}\biggr)\biggr]
\label{eqn:covmat}
\end{equation}
where $\btheta\equiv (\theta_x,\theta_y)$ is the angular position of star $i$
relative to the center, and $N_r$ is the number of radial-velocity 
measurements.
However, for simplicity, I will consider stars distributed uniformly over
a circular area of radius $\Delta\theta$.  In this case, the error in
each component of $\nabla v_r$ (or equivalently $\bV_\perp$, since
$\nabla u_r$ is assumed known) is
\begin{equation}
\sigma(V_\perp) = \sigma_\nabla =
c_{kk}^{1/2} = \biggl({2\over N_r}\biggr)^{1/2}
{\sigma_v\over\Delta\theta}.
\label{eqn:sigmavperp}
\end{equation}
See also Narayanan \& Gould (1999).

	Suppose now that the proper motion $\bmu$ of the system is measured
with error $\sigma_\mu$
The distance and distance error are then,
\begin{equation}
d = {V_\perp\over \mu},\qquad
\biggl({\sigma_d\over d}\biggr)^2 = 
\biggl({\sigma_\nabla\over V_\perp}\biggr)^2 + 
\biggl({\sigma_\mu\over \mu}\biggr)^2
\label{eqn:sigmad}
\end{equation}
That is, the fractional distance error is limited by the larger of the
errors in the transverse velocity and the proper motion.

\section{Application to the LMC}

\subsection{Naive}

	Substituting values appropriate for the LMC into equation
(\ref{eqn:sigmavperp}), I obtain
\begin{equation}
{\sigma_\nabla\over V_\perp} = 1.5\%\biggl({N_r\over 10^4}\biggr)^{1/2}
{\sigma_v\over 24\,\kms}
\biggl({V_\perp\over 325\,\kms}\biggr)^{-1}
\biggl({\Delta\theta\over 4^\circ}\biggr)^{-1}
\label{eqn:lmceval}
\end{equation}
where I have chosen a dispersion characteristic of carbon stars
(Cowley \& Hartwick 1991) and the estimate of the transverse velocity from
the proper-motion measurement of Jones, Klemola, \& Lin (1994).
Hence, good statistical precision is
possible provided that a large sample of stars is available.  Note that
the measurement errors are not important provided that they are well
below dispersion.  Since $\sigma_v\sim 5\,\kms$ errors are not 
difficult to achieve for LMC carbon stars, it is feasible to obtain a very
large sample such as is envisaged in equation (\ref{eqn:lmceval}).

	While the proper motion of the LMC is only crudely known today
(Jones et al.\ 1994; Kroupa \& Bastian 1997), it could be measured 
to very high precision with any of a number of proposed astrometry 
satellites including FAME, SIM, and GAIA.  For definiteness, I will
focus on the capabilities of FAME which has the earliest possibility of
launch.  I find
from the USNO-A2.0 catalog (Monet 1998), that there are a total of
21,900 stars with $13<V<15$ within $\Delta\theta=4^\circ$ of the center
of the LMC at $(l,b) = (280.5,-32.9)$, where I estimate $V=(B+R)/2$.
Of these, about 13,300 are foreground Galactic stars as judged from counts
in three similar circles at $(l,b) = (280.5,+32.9$), and
$(l,b)=(79.5,\pm 32.9)$.  This leaves
$N_\perp\sim$ 8,600 stars in the LMC.  The dispersions of
LMC stars in the transverse directions are unknown, but based on what is
known of disk kinematics in the Galaxy, it is plausible to assume that
they are $\sim 50\%$ higher than the vertical dispersion, or 
$\sigma_\perp\sim 35\,\kms$.  
Hence, if the proper motions of these stars could be measured to better
than $\sigma_\perp/\dlmc\sim 150\,\mu\rm as\,yr^{-1}$, and if the internal
systemic motions $\bu$ are again assumed known (see \S\ 4), then
the precision of the LMC proper motion would be given by
\begin{equation}
{\sigma_\mu\over \mu} = N_\perp^{-1/2}\,{\sigma_\perp\over V_\perp} =
0.1\%\biggl({N_\perp\over 8600}\biggr)^{-1/2}
{\sigma_\perp\over 35\,\kms}
\biggl({V_\perp\over 325\,\kms}\biggr)^{-1},
\label{eqn:sigmuovermu}
\end{equation}
where $N_\perp$ is the number of proper-motion measurements.
In fact, FAME probably cannot achieve quite this precision at $V=15$,
but should come within a factor of 2 (Horner et al.\ 1998)
and so easily achieve $\sigma_\mu/\mu\la 1\%$
or $\sigma_\mu\la 10\,\mu\rm as\,yr^{-1}$.  The present rotational
precision of the extra-galactic reference frame is 
$\sigma_\mu\sim 5\,\mu\rm as\,yr^{-1}$.  However, the FAME astrometric
frame will probably be accurate only to within 
$\sigma_\mu\sim 25\,\mu\rm as\,yr^{-1}$ (assuming 100 QSOs with 
$V\la 15$ and hence with
mean proper motion errors of $250\,\mu\rm as\,yr^{-1}$).  
The FAME frame will be fixed substantially better by SIM.
In brief, the proper-motion measurement error can probably be reduced to about
2\% with FAME alone and substantially less by combining FAME and SIM.

\subsection{Degeneracy}

	However, the potentially fatal flaw in this method is that $\bu$
is {\it not} known (as has been assumed so far) and must be determined from
the same kinematic data that are used to derive the distance measurement.
As is well known from the classical application of the radial-velocity
gradient method to the Hyades (Detweiler et al.\ 1984; Gunn et al.\ 1988) and
the Pleiades (Narayanan \& Gould 1999), if the cluster were undergoing
solid-body rotation 
$\bu = \bOmega\times\br$, this would produce a radial-velocity gradient
\begin{equation}
\nabla u_r = (\bn_0\times \bOmega)d.
\label{eqn:spurious}
\end{equation}
Here $\br$ is the 3-space position of a star relative to the cluster center.
This gradient is indistinguishable from the gradient produced by a transverse
velocity and so, if unrecognized, would corrupt the distance measurement
given by equation (\ref{eqn:sigmad}).  In the case of clusters, one
normally simply assumes that the cluster is not rotating.  However, one can
check this assumption by comparing the directions of the radial velocity
gradient and the proper motion.  If these differ, the cause might be
rotation (or systematic errors).  If they are the same, then either the
cluster is not rotating, or its rotation happens to be perfectly aligned with
its proper motion (within statistical errors).

	The situation is similar for the LMC but is somewhat more complicated
because the LMC {\it is} rotating.  While the rotation is not solid body,
it can be reasonably approximated as such in the inner $2^\circ\hskip-2pt .5$.  
To the extent the rotation is solid-body, one measures a gradient
\begin{equation}
\nabla v_r = \bV_\perp + \bOmega_\times \dlmc, \qquad
\bOmega_\times \equiv \bn_0\times \bOmega,
\label{eqn:nablavr}
\end{equation}
and from this measurement alone, has no idea how to separate the
two components.  If, for example, one ignored the transverse motion, one
would interpret the gradient as due entirely to rotation and would therefore
misjudge the amplitude of rotation.  One would misjudge 
its orientation as well to the
extent that $\bV_\perp$ does not happen to lie parallel to
$\bOmega_\times$.

\subsection{Breaking the Degeneracy}

	However, for a disk rotating about its axis of symmetry,
$\bOmega_\times$ should be aligned with the apparent major axis of the
system, i.e., the photometric line of nodes.  This provides some
information with which to break the degeneracy.

	These effects were first investigated when Feitzinger et al.\
(1977) reanalyzed earlier kinematic data.  They noted that the kinematic
line of nodes (locus of extrema in radial velocity)
was displaced by $\sim 20^\circ$ from the photometric line of nodes 
(major axis of the surface-brightness profile) at 
position angle $\phi_{\rm ph}=-10^\circ$.  They assumed that
this displacement was caused by transverse motion in the direction
$\phi_\mu=110^\circ$ (i.e., the direction of the Magellanic stream)
and then solved for the amplitude of this motion $V_\perp\sim 275\,\kms$.
Subsequently, several other workers applied a similar
procedure to various stellar samples and obtained various results
(Rohlfs et al.\ 1984; Meatheringham et al.\ 1988; Hughes, Wood, \& Reid 1991).
Note that this approach to breaking the degeneracy
requires {\it two} pieces of information in addition
to the kinematic data: first the position angle of the photometric line
of nodes $\phi_{\rm ph}$, and second the direction of LMC motion $\phi_\mu$.

	However, if the proper motion is measured 
(which is necessary in any case to determine the distance through eq.\ 
\ref{eqn:sigmad}), one already knows $\phi_\mu$.  From equation
(\ref{eqn:nablavr}), the three vectors
$\nabla v_r$, $\bV_\perp$, and $\bOmega_\times\dlmc$
form a triangle, so by the law of sines,
\begin{equation}
V_\perp = {\sin(\phi_\nabla-\phi_{\rm ph})\over\sin(\phi_\mu-\phi_{\rm ph})}
|\nabla v_r|,
\label{eqn:vperpeval}
\end{equation}
where $\phi_\nabla$ is the observed position angle of the kinematic line 
of nodes.  The quantities on the right-hand side of equation
(\ref{eqn:vperpeval}) are all observables.  Assuming that the
errors in the measurements of $\bmu$ and $\nabla v_r$ are isotropic, so
that $\sigma(\phi_\mu)=\sigma_\mu/\mu$ and 
$\sigma(\phi_\nabla)= \sigma_\nabla/|\nabla v_r|$, one can
evaluate the error in $\dlmc=V_\perp/\mu$ by taking the derivatives of
equation (\ref{eqn:vperpeval}) with respect to the various parameters.
I find,
\begin{equation}
\biggl({\sigma_d\over d}\biggr)^2 = \csc^2(\phi_\mu-\phi_{\rm ph})\biggl[
\biggl({\sigma_\nabla\over V_\perp}\biggr)^2 + 
\biggl({\sigma_\mu\over \mu}\biggr)^2\biggr]
+ \biggl[{\sin(\phi_\mu-\phi_\nabla)\over
\sin(\phi_\nabla-\phi_{\rm ph})\sin(\phi_\mu-\phi_{\rm ph})}\biggr]^2
\sigma_{\rm ph}^2,
\label{eqn:sigmad2}
\end{equation}
where $\sigma_{\rm ph}$ is the error in the determination of $\phi_{\rm ph}$.

Equation
(\ref{eqn:sigmad2}) differs from its naive relative, equation
(\ref{eqn:sigmad}), in two ways.  First, the entire error in equation
(\ref{eqn:sigmad}) is now multiplied by a factor $\csc(\phi_\mu-\phi_{\rm ph})$.
Second, there is a new term which is related to the uncertainty
in the photometric position angle.  To understand the importance of these
changes, I first introduce representative values of the parameters.  I 
choose $\phi_\mu= 97^\circ$ from the proper motion meausrement of
Jones et al.\ (1994), $\Omega_\times=12\,\kms\,\rm kpc^{-1}$, and
$V_\perp=325\,\kms$.  Together, these imply $\phi_\nabla=14^\circ$, thus
$(\phi_\mu-\phi_{\rm ph})=107^\circ$, 
$(\phi_\nabla-\phi_{\rm ph})=24^\circ$, and
$(\phi_\mu-\phi_\nabla)=83^\circ$.  The fact that $\phi_\mu$
and $\phi_{\rm ph}$ are almost at right angles implies that the
$\csc^2(\phi_\mu-\phi_{\rm ph})$ term in equation (\ref{eqn:sigmad2}) is
essentially unity.  However, since $\Omega_\times \dlmc\gg V_\perp$, the
radial-velocity gradient due to $\bV_\perp$ is a relatively minor perturbation
on the gradient due to internal motion, and so $\phi_\nabla$ is not much 
different from $\phi_{\rm ph}$.  Hence, the factor
$\sin(\phi_\nabla-\phi_{\rm ph})=0.41$ in the denominator of the last
term is relatively small.  This means that $\phi_{\rm ph}$ must be measured
quite accurately if one wants a precise measurement of the effect of
the transverse velocity.  Specifically, the last term in equation
(\ref{eqn:sigmad2}) is $(2.55\sigma_{\rm ph})^2$.  At distances from the
center $\ga 2^\circ\hskip-2pt .5$, the rotation curve tends to flatten,
and so $\bV_\perp$ becomes a larger relative perturbation
causing $(\phi_\nabla-\phi_{\rm ph})$ to grow and thus making the measurement
somewhat easier.  Nevertheless, imprecise knowledge of $\phi_{\rm ph}$ is 
likely to be a major limitation of the method.

\section{Measurement of $\phi_{\rm ph}$}

	To achieve
2\% precision in $\sigma_d/d$ (which generally seems feasible from
the standpoint of the $\nabla v_r$ and $\bmu$ measurements) would require
measuring the position angle to $\sigma_{\rm ph}\sim 0^\circ \hskip-2pt .4$,
or 0.008 radians.
It is difficult to believe that this can be achieved using surface photometry
alone.  Recall, that one is not actually interested in the best
fit to the major axis of the isophotes.  Rather, one wants to know the
position angle of the line that crosses the plane of the sky.  Certainly
star formation, dust, etc corrupt the surface-brightness profile too much
to extract information at this level of precision.  
It should be possible
to make a more accurate assessment of $\phi_{\rm ph}$ using star counts 
particularly of clump giants.  Using the method of Gould (1995a) one may
show that this technique can determine $\phi_{\rm ph}$ with precision
\begin{equation}
\sigma_{\rm ph} =
\biggl[{N_{\rm cg}\over 8}\biggl\langle\biggl({d\ln F\over d\ln R}\biggr)^2
\biggr\rangle\biggr]^{-1/2}{\cos i\over \sin^2 i}\sim 0^\circ \hskip-2pt .2
\biggl({N_{\rm cg}\over 10^6}\biggr)^{-1/2}
\label{eqn:photdet}
\end{equation}
where $N_{\rm cg}$ is the number of clump giants, $i$ is the inclination
of the disk, $F(R)$ is the (assumed axially symmetric) radial profile of
the LMC disk, and where I have assumed $i=30^\circ$, and
$\langle(d\ln F/d\ln R)^2\rangle=6$, which is valid for an exponential
disk.

	However, clump giants provide another, independent route to the
measurement of the position angle.  Clump giants have a dispersion in $I$
band of only $\sigma_{\rm cg} = 0.15$ mag (Udalski et al.\ 1998).  
The stars on the near side should therefore be brighter than those on the
far side by a significant fraction of this dispersion.  Averaging this
effect over the whole disk, I find that $\sigma_{\rm ph}$ can be determined
to a precision
\begin{equation}
\sigma_{\rm ph} = \biggl({N_{\rm cg}\over 2}\biggr)^{-1/2}{\ln 10\over 5}\,
{\dlmc\over\langle R^2\rangle^{1/2}}\,\csc i\,\sigma_{\rm cg}
\sim 0^\circ \hskip-2pt .15\,\biggl({N_{\rm cg}\over 10^6}\biggr)^{-1/2}
\label{eqn:photdet2}
\end{equation}
where I have assumed an exponential scale length of $\alpha=1^\circ \hskip-2pt
.7$ (Feitzinger et al.\ 1977), 
so that $\langle R^2\rangle/d_{\rm LMC}^2 = 6\alpha^2$.
The challenges to actually carrying out such a measurement would be 
formidable.  Just maintaining a constant photometric zero point 
at the level of $\sigma_{\rm cg}/N_{\rm cg}^{1/2}\sim 10^{-4}$ mag over fields
separated by $\sim 10^\circ$ would be difficult.  In addition, one would
have to correct for differential reddening, probably from
the clump giant colors, but to do so would require an accurate estimate
of $E(V-I)/A_I$.  This could be made empirically by looking at the
correlation between $V-I$ and $I$ at fixed position but might not be easy.

	In principle, it is also possible to measure $\phi_{\rm ph}$ from
the transverse velocity field measured from the proper motions.  In
practice, however, the errors in this determination are too large for it
to be useful.  Note that the internal transverse motions do not increase
the uncertainty in $\bmu$.  The uncertainty in the transverse velocity field at
any particular point is much smaller than either the dispersion or the
measurement error, and there is no uncertainty in the mean internal motion
averaged over the whole population: the mean internal motion is zero.

\section{Discussion}

	I have outlined how the radial-velocity gradient method could be
applied to measure $\dlmc$ with statistical errors of 2\% or less.  Of
course, as in most distance measurements beyond the solar neighborhood, 
the largest potential source of errors is systematics.  Examples of effects
that would generate such systematic errors are non-circular motions and/or
warps in the LMC disk and contamination by material along the line of sight.
For example, Weinberg (1999) has recently shown that resonant interactions
between the Milky Way and the LMC can profoundly disturb the LMC disk.

	However, given the mass of data required to make the measurements,
it should be possible to conduct many tests for systematics.  For example,
non-circular motions would affect both the radial-velocity gradient
and the orientation of the photometric line of nodes.  The latter would
have a larger impact on the distance simply because its coefficient in
equation (\ref{eqn:sigmad2}) is $\sim 2.4$ times larger.  Such motions
should be revealed in the comparison of the clump-giant star-count and
photometric methods for measuring $\phi_{\rm ph}$: the star-count method
would be affected by non-circular motions while the photometric method
would not.  Both warps and non-circular motions could be tested by
comparing the radial-velocity field with the transverse-velocity field obtained
from proper motions.  Similarly, it is possible to search the radial
velocities for evidence of unassociated material along the LMC line of sight
(Graff et al.\ 1999).

	While an all-out serach for systematic effects probably requires the
full data set, substantial initial investitgations can be made with
existing phtometric catalogs or with radial velocity studies now
underway (e.g., Suntzeff, Schommer, \& Hardy 1999).  

	I have estimated that FAME will obtain 8600 proper motions with
a mean precision of $250\,\mu\rm as\,yr^{-1}$.  If FAME is not launched, what
are the prospects for matching this performance?  Clearly GAIA, which is
also a survey mission but with much higher precision and fainter magnitude
limits could easily meet this standard.  However, given its later launch data
and larger analysis time, GAIA would require an additional decade to produce
results.  SIM certainly has the capability to make these measurements,
but whether it would make so extensive a survey is open to question.
Recall that the proper-motion measurements need only be a factor of a few
better than the internal dispersion ($\sim 150\mu\rm as\,yr^{-1}$).  For
$V\sim 15$ stars, SIM could do an order of magnitude better than this
in 1 minute.
Allowing another minute for pointing and assuming a total of 4 position
measurements per star, 8600 proper-motion measurements would require
about 1000 hours.  From equation (\ref{eqn:sigmuovermu})
only $\sim 100$ stars would be needed to measure $\dlmc$ to $\sim 1\%$, and
this could be done in only about 10 hours.  In this case, however,
one would lose much of the ability to check for systematics from a 
comparison of the radial-velocity and proper-motion fields.  In brief,
FAME is the instrument of choice to make the proper-motion measurement.

{\bf Acknowledgements}: I thank David Graff for stimulating discussions
and David Weinberg thoughtful comments on the manuscript.  
This research was supported in part by grant AST 97-27520 from the NSF,
and in part by grant NAG5-3111 from NASA.

\newpage



\clearpage 

\begin{references} 
\reference{} Cowley, A.P., \& Hartwick, F.D.A.\ 1991, \apj, 373, 80
\reference{} Detweiler, H.L., Yoss, K.M., Radick, R.R. \& Becker, S.A., 1984, 
\aj, 89, 1038
\reference{} Feast, M.W., \& Catchpole, R.M.\ 1997, \mnras, 286, L1
\reference{} Feitzinger, J.V., Issertedt, J., \& Schmidt-Kaler, T.\ 1977,
\aap, 57, 265
\reference{} Gould, A.\ 1995a, \apj, 440, 510
\reference{} Gould, A.\ 1995b, \apj, 452, 189
\reference{ga} Gould, A., Bahcall, J.N., \& Flynn, C.\ 1997, \apj, 482, 913
\reference{ga} Gould, A., Flynn, C., \& Bahcall, J.N.\ 1998, \apj, 503, 798
\reference{} Gould, A., \& Popowski, P.\ 1998, \apj, 508, 844
\reference{} Gould, A., \& Uza, O.\  1998, \apj, 494, 118
\reference{} Graff, D., Gould, A.,
Suntzeff, N.B., Schommer, R.A. \& Hardy, E.\ 1999, in preparation
\reference{} Gratton, R.\ G, Pecci, F.\ F., Carretta, E., Clementini, G., 
Corsi, C., E., \& Lattanzi, M.\ 1997, \apj, 491, 749
\reference{} Gunn, J.E., Griffin, R.F., Griffin, R.E.M. \& Zimmerman, B.A., 
1988, \aj, 96, 198 
\reference{} Horner, S.D., et al.\ 1998, BAAS, 193, 12.06
\reference{} Hughes, S.M.G., Wood, P.R., \& Reid, N.\ 1991, \aj, 101, 1304
\reference{} Jones, B.F., Klemola, A.R., \& Lin, D.N.C.\ 1994, \aj,
107, 1333
\reference{} Kroupa, P., \& Bastian, U., New Astronomy, 2, 77
\reference{} Madore, B.F., et al.\ 1999, ApJ, 515, 29
\reference{} Meatheringham, S.J., Schwarz, M.P., \& Murray, J.D.\ 1977, \apj,
217, L5
\reference{monet} Monet, D.\ 1998, BAAS, 193, 120.03
\reference{} Narayanan, V.K., \& Gould, A.\ 1999, \apj, 523, 000
\reference{} Panagia, N.\ 1998, New Views of the Magellanic Clouds
IAU Symposium 190
\reference{} Panagia, N., Gilmozzi, R., Macchetto, F., Adorf, H.-M. \&
Kirshner, R.\ P.\ 1991, \apj, 380, L23
\reference{} Popowski, P., \& Gould, A.\ 1999, Post-Hipparcos Cosmic Candles,
A.\ Heck \& F.\ Caputo, eds., p.\ 53, (Kluwer: Dordrecht)
\reference{} Reid, I.\ N.\ 1997, \aj, 114, 161
\reference{} Rohlfs, A., Kreitschmann, J., Seigman, B.C., \& Feitzinger, J.V.\
1984, \aap, 137, 343
\reference{} Saha, A., Sandage, A., Tammann, G.A., Labhardt, L., 
Macchetto, F.D., \& Panagia, N.\ 1999, ApJ, in press
\reference{} Sonneborn, G., Claes, F., Lundqvist, P.,
Cassatella, A., Gilmozzi, R., Kirshner, R.P., Panagia, N.,
\& Wamsteker, W., \apj, 477, 848
\reference{}Suntzeff, N.B., Schommer, R.A. \& Hardy, E.\ 1999, in preparation
\reference{} Udalski, A., Szyma\'nski, Kubiak, M., Pietrzy\'nski, G.,
Wo\'zniak, P., \& \.Zebru\'n, K.\ 1998, Acta Astronomica, 48, 1
\reference{} Walker, A.\ 1992, ApJ, 390, L81
\reference{weinberg} Weinberg, M.\ 1999, preprint (astro-ph/9905305)
\end{references}
\end{document}